\documentstyle[]{article}
\begin{document}
\begin{center}
{{\bf \huge Quantum Critical Phenomena, Entanglement Entropy and Hubbard Model in 1d with the Boundary Site with a Negative Chemical
Potential -p and the Hubbard Coupling U Positive
}

\large
\vskip1cm
O. Hudak
\vskip1cm

Department of Physics, Pedagogical Faculty, Catholic University,  Nam. Andreja Hlinku 56/1, SK - 034 01 Ruzomberok \\ and \\
Department of Aerodynamics and Simulations, Faculty of Aeronautics, Technical University Kosice, Rampova 7,  SK - 040 01 Kosice
}

\end{center}
\newpage 
\section*{Abstract}
Recently the ground state and some excited states of the half-filled case of the 1d Hubbard model were discussed for an open chain with L sites. Authors considered the case when the boundary site has a negative chemical
potential -p and the Hubbard coupling U is positive. They have shown by an analytic method that when p is larger than the transfer integral some
of the ground-state solutions of the Bethe ansatz equations become
complex-valued. They have found that there is a "surface phase transition" at some critical value $p_c$; when $p < p_c$ all the charge excitations have the gap for this case,  while there exists a massless charge mode when $p > p_c$ . To find whether this "surface phase transition" is of the first order or of the second order we have used the entanglement entropy concept. The entropy and its derivative has a discontinuity there, so this transition is of the first order.

\newpage 
\section{Introduction}
Quantum critical phenomena are phenomena where the transition is driven by quantum effects in the system of many electrons and bosons. Quantum phase transition with nonstandard behaviour was described in \cite{VJ}, it is a phase transition and a critical behaviour deviating from the standard scheme based on the mean field theory renormalising only the mass of the critical excitations and with a perturbative scaling renormalisation of the interaction strength. 

Criticality in 1d-Hubbard model was studied recently using entanglement scaling \cite{LJ}. Entanglement is a quantity which  characterises nonlocal correlations in quantum systems. There is a connection between the entanglement of a many particle system and the quantum critical phenomena. Discontinuity in the entanglement  derivative seems to be \cite{LJ} associated with the first order QPT, singularity with the second order. Entanglement describes how the associated nonlocal quantum correlations influence the critical behaviour of a quantum phase transition. As noted by the authors \cite{LJ} to what extent their results can be generalised to other quantum systems is yet to be answered. In this paper we will discus entanglement for the 1d-Hubbard model with the boundary site with a negative chemical potential -p and the Hubbard coupling U positive for a finite number of atoms in the chain. In general we have found that there is a 1-st order "surface phase transition" phenomena present.

\section{Quantum Critical Phenomena and Entanglement.}

The Hubbard model displays many features which are close to real properties of materials with electron-electron interactions. Some rigorous results concerning this model are described in \cite{L}. These results and the entanglement concept mentioned in the preceding section enable to study the quantum critical phenomena in 1d-Hubbard model. Let us describe them shortly.

Exact solution of the Hubbard model with boundary hoppings and fields was found in \cite{KE}. The Bethe ansatz method was used. It was found that for certain values of the boundary hoping integrals and fields the ground state contains boundary bound states. While in \cite{LJ} the authors assume periodic boundary conditions, in \cite{KE} the boundary conditions are such that periodicity is broken. The Hamiltonian in \cite{LJ} is transnational invariant and conserves particle number and the z component of the total spin. When the boundary fields and hoppings are present then this symmetry is broken. However boundary fields and hoppings are expected to play no role in the limit of large lattice, $L \rightarrow \infty$. The reduced density matrix $\rho_{A}$ for a single site A is diagonal in the chosen basis for periodic  boundary conditions. For boundary fields and hoppings or other boundary effects present we expect that it is diagonal in the limit of large lattice, $L \rightarrow \infty$.

The single-site entanglement entropy of the ground state 
$| \psi_{0}> $ can be written \cite{LJ}:
\begin{equation}
\label{1}
E = - w_{0} log_{2} (w_{0}) - w_{u} log_{2} (w_{u}) - w_{d} log_{2} (w_{d}) - w_{2} log_{2} (w_{2})
\end{equation}
where
\begin{equation}
\label{2}
w_{2} = <n_{ju}n_{jd}> 
\end{equation}
\[ w_{\alpha} = <n_{j \alpha}>  - w_{2}, \] 
\[ w_{0} = 1 - w_{u} - w_{d} - w_{2}.\] 
where $\alpha = u, d$ are spin up and down indices.

In the limit of large $ |u| \equiv |\frac{U}{4t}| >> 0$ the Hellman-Feynman theorem can be used, $<\frac{\delta H}{\delta u}>_{0} = 
\frac{ \delta E_{0}}{ \delta u} $. The Bethe ansatz result for the ground state energy gives energy proportional to L, the number of sites in the lattice. There is present also a contribution to this energy from the surface which is of the order of $L^{0}$, and contributions of the order of $L^{-1}$, see in \cite{AS}.

\section{Hubbard Model in 1d with the Boundary Site with a Negative Chemical Potential -p and the Hubbard Coupling U Positive}

The surface energy for the boundary fields was calculated exactly for the XXZ chain by the Bethe ansatz in \cite{KS}. The energy is an analytic function of the field, where $h=h_{1}=-h_{2}$ for the case in which there are opposite oriented fields $h_{1}$ and $-h_{2}$ on boundaries. While there are identified "critical" fields $h_{1c}$ and $h_{2c}$ as fields below which and above which respectively (in absolute values) there are present 1-string boundary states in the chain, for fields with values in between $h_{1c}$ and $h_{2c}$ there are no boundary strings. Then using the entanglement we find that there is no discontinuity or divergence in its derivative for this case. So there is no quantum phase transition changing boundary fields and thus the ground states between the boundary-string state and the state without the boundary string state. The XXZ model is related to the Hubbard model by a canonical transformation. Thus we can expect that at the spin sector of the solutions of the Hubbard model with boundary fields and hoppings there is no "surface phase transition".

Let us now concentrate on the charge sector. In \cite{DYK} the Hubbard model is discussed for the ground state and some excited states of the half-filled case for an open chain with L sites. Only one of the boundary sites has a different value of chemical potential in this case. Authors \cite{DYK} considered the case when the boundary site has a negative chemical potential -p and the Hubbard coupling U is positive. They have shown by an analytic method that when p is larger than the transfer integral some of the ground-state solutions of the Bethe ansatz equations become complex-valued. They have found, among other findings, that there is a "surface phase transition" at some critical value $p_{3c}$; when $p < p_{3c} $ all the charge excitations have the gap for this case,  while there exists a massless charge mode when $p > p_{3c} $ . The authors did not discussed whether the "surface phase transition" is of the first or of the second order in the sense discussed above for the entanglement.

To find whether this "surface phase transition" is of the first order or of the second order we can use the entanglement entropy described above. We have found using the results of \cite{DYK} that this entropy has a discontinuity there for zero magnetic field. Consequently the difference of derivatives of entanglement entropy with respect to the chemical potential is nonzero at $p = p_{3c} $ calculating it in the corresponding limiting procedure from below and from above $p = p_{3c} $. This difference of derivatives is proportional to a (charge) susceptibility and to a coefficient which is finite with exception of two points $0$ and $\frac{1}{2}$ for $w_{2}$. Thus in general there is a 1-st order "surface phase transition" phenomena present. The value of the quantity $w_{2}$ for $p = p_{3c} $ is in general different from the values corresponding to the mentioned two points for larger values of u when correlations of fluctuations of the  
number of electrons with spin up and with spin down is nonzero. For a finite chain with L finite we have for $p > p_{3c} $ a hole at the boundary site 1, and for $p < p_{3c} $ there is an electron at this boundary site 1 \cite{DYK}. As we can see there is no phase transition in the thermodynamic limit. That is the reason why we speak about the "surface phase transition" in this model. Corrections to entanglement entropy are not logarithmic here, however again reflect a change in the number of local states accessible to the system at the transition as it was described above.

\section*{Acknowledgement}
This paper is a part of study done within the VEGA project 1/3042/06. The author would like to express his sincere thanks to Peter  Prešnajder from FMFI UK  Bratislava for his kind support and to Vaclav Janis from the Institute of Physics AS CR Prague for interesting discussions about entanglement, 1d-Hubbard model and quantum critical phenomena and for kind hospitality.

%

\end{document}